\documentclass[comsoc, conference,10pt]{IEEEtran}

\usepackage{cite}
\usepackage{physics}

\usepackage{bbm}

\usepackage{boldline}

\usepackage{multirow}
\usepackage{graphicx}
\usepackage{textcomp}
\usepackage{subcaption}
\def\BibTeX{{\rm B\kern-.05em{\sc i\kern-.025em b}\kern-.08em
    T\kern-.1667em\lower.7ex\hbox{E}\kern-.125emX}}

\usepackage{listings}
\usepackage{tikz}

\usetikzlibrary{positioning,shapes.gates.logic.US} 

\newcommand{\expnum}[1]{#1}

\usepackage{xcolor, soul}

\definecolor{bananamania}{rgb}{0.98, 0.91, 0.71}
\sethlcolor{bananamania}

\newcommand{\ry}[1]{\gate{R_y({#1})}}
\newcommand{\rypi}[1]{\gate{R_y(\frac{\pi}{{#1}})}}

\usepackage{amsthm}

\usepackage[titlenumbered,ruled,linesnumbered,noend]{algorithm2e}
\usepackage{cleveref}

\usepackage{float}

\usepackage[]{qcircuit}
\usepackage{amsmath,amssymb,amsfonts}

\title{\huge Quantum State Preparation Using an Exact\\ CNOT Synthesis Formulation}

\author{
    \IEEEauthorblockN{Hanyu Wang$^{1}$, Bochen Tan$^{2}$, Jason Cong$^{2}$, and Giovanni De Micheli$^{3}$}
    \IEEEauthorblockA{$^{1}$\textit{ETH, Zurich, Swizterland},\, $^{2}$\textit{UCLA, California, United States},\, $^{3}$\textit{EPFL, Lausanne, Swizterland}}
}

\begin{document}
\makeatletter
\patchcmd{\@maketitle}
  {\addvspace{0.5\baselineskip}\egroup}
  {\addvspace{-0.8\baselineskip}\egroup}
  {}
  {}
\makeatother

\renewcommand{\baselinestretch}{0.95}

\maketitle

\begin{abstract}
Minimizing the use of CNOT gates in quantum state preparation is a crucial step in quantum compilation, as they introduce coupling constraints and more noise than single-qubit gates. Reducing the number of CNOT gates can lead to more efficient and accurate quantum computations. However, the lack of compatibility to model superposition and entanglement challenges the scalability and optimality of CNOT optimization algorithms on classical computers. In this paper, we propose an effective state preparation algorithm using an exact CNOT synthesis formulation. Our method represents a milestone as the first design automation algorithm to surpass manual design, reducing the best CNOT numbers to prepare a Dicke state by 2$\times$. For general states with up to 20 qubits, our method reduces the CNOT number by 9\% and 32\% for dense and sparse states, on average, compared to the latest algorithms.
\end{abstract}

\section{Introduction}
\emph{Quantum states preparation}~(QSP) is essential for compiling quantum algorithms~\cite{childs2000finding}, implementing quantum communications~\cite{ben2005fast}, studying quantum metrology~\cite{toth2012multipartite}, and experimenting with quantum entanglement~\cite{horodecki2009quantum}. For decades, researchers have based QSP circuit designs on mathematical derivation~\cite{cruz2019efficient, bartschi2019deterministic, mukherjee2020preparing, aktar2022divide} and created them manually. Although efficient, the derived properties are specialized to a certain class of highly symmetric states, e.g., GHZ states~\cite{greenberger1989going}, W states~\cite{dur2000three}, and Dicke states~\cite{dicke1954coherence}, and cannot be generalized. 

Limitations of manual designs point us to the need to develop design automation algorithms. Recent works propose Boolean methods utilizing decision diagrams to prepare general $n$-qubit states using $\mathcal{O}(2^n)$ CNOT gates~\cite{araujo2021divide, mozafari2019preparation, niemann2016logic}. By leveraging sparsity, the latest studies improve the efficiency and develop algorithms to prepare $n$-qubit with $m$ nonzero amplitudes using $\mathcal{O}(mn)$ CNOT gates~\cite{gleinig2021efficient, mozafari2022efficient, malvetti2021quantum}. However, these methods sacrifice optimality for efficiency, consuming more CNOT gates than manual designs.

The challenges of solving QSP on classical computers are the complexities of modeling superposition and entanglement~\cite{nielsen2010quantum}. Indeed, while classical computers store binary information, compute Boolean operators, and retrieve a single binary output, quantum computing processes high-dimensional state vectors with complex amplitudes that evolve with matrix multiplications. With superposition, the dimension of state vectors grows exponentially with the number of qubits. Therefore, only certain families of quantum states can be efficiently encoded using classical bits~\cite{aaronson2004multilinear, raz2009multi}. Besides, because of entanglement, the qubits in the states are inseparable. As a result, developing an effective divide-and-conquer approach for QSP problems is difficult.

In this paper, we propose an exact CNOT synthesis formulation for QSP. Our method encodes quantum states and gates on a graph and formulates QSP as a shortest path problem. Given an arbitrary quantum state and a library of available gates, our formulation provides full visibility of the solution space and is guaranteed to find the optimal circuit with minimal CNOT gates. For general states with up to $20$ qubits, our method reduces the CNOT number by $9\%$ and $32\%$ for dense and sparse states, on average, compared to the latest algorithms. Besides, our method represents a milestone as the first design automation algorithm to surpass manual design, reducing the best CNOT numbers to prepare a Dicke state by $2\times$.

In the rest of the paper, we present the background in \Cref{sec:background} and show an example to motivate our work in \Cref{sec:motivation}. Then, we illustrate our shortest path problem formulation for QSP in \Cref{sec:formulation} and introduce our specialized shortest path algorithm in \Cref{sec:algorithm}. In \Cref{sec:evaluation}, we demonstrate experimental results and evaluate our method.

\section{Background}\label{sec:background}
In this section, we provide background for quantum state preparation. For clarity and space constraints, we refer readers to established sources for formal definitions of notations~\cite{nielsen2010quantum} and quantum gates\cite{barenco1995elementary}.

\subsection{Quantum States and Quantum Gates}
We express the $n$-qubit \emph{quantum state} as a linear combination of $2^n$ orthonormal basis vectors.
\begin{equation*}
    \ket{\psi} = \sum_{x\in S(\psi)} c_x \ket{x},\,\text{and}\, \sum_{x\in S(\psi)} |c_x|^2 = 1,
\end{equation*}
where $\ket{x}\in\{0,1\}^n$ is the \emph{basis state}, $c_x\in\mathbb{C}$ are \emph{amplitudes} whose norm indicates the probability of observing $\ket{x}$, and $S(\psi)$ is the \emph{index set}, which represents the set of basis with nonzero amplitudes. The \emph{cardinality} of a state $\psi$ is the number of elements in its index set, $|S(\psi)|$.

\emph{Quantum gates}, or \emph{operators}, denoted by $U$, are unitary matrices representing transitions between quantum states. $\mathcal{U}(2^n)$ represents the set of all $n$-qubit gates. This paper studies states with \textbf{real} amplitudes and restricts state transitions in the \textbf{X-Z plane}. Therefore, all single-qubit unitary matrices, $U\in\mathcal{U}(2)$, are Y rotations, $\text{R}_y$, and all multi-qubit operators can be decomposed into gates in $\{\text{CNOT}, \text{R}_y\}$.
\begin{equation}\label{eqn:u-ry}
    U = \begin{pmatrix}
        a & -b \\
        b & a 
    \end{pmatrix} = \begin{pmatrix}
        \cos\frac{\theta}{2} & \sin\frac{\theta}{2} \\
        -\sin\frac{\theta}{2} & \cos\frac{\theta}{2}
    \end{pmatrix} = \text{R}_y(\theta),
\end{equation}
where $a$ and $b$ are real numbers satisfying $a^2+b^2=1$ and $\theta$ is a rotation angle that satisfies $\theta = -2\cdot\arctan\frac{b}{a}$.

We define \emph{CNOT cost} of an operator $U$ as the number of CNOT gates to decompose $U$. Quantum gates involved in this paper and their costs are listed in \Cref{tab:gates}, including \emph{Y rotations}~(R$_y$), CNOT, and \emph{multi-controlled Y rotations}~(MCR$_y$) gates. Note that the CNOT cost depends on the decomposition algorithm and the number of ancillary qubits~\cite{maslov2016advantages, mottonen2004quantum, saeedi2013linear}. This paper assumes an MCR$_y$ gate with $n$ control qubit has a CNOT cost of $2^n$. 

\begin{table}[t]
    \caption{\small Selected quantum gates with their CNOT costs.}
    \label{tab:gates}
\renewcommand{\arraystretch}{1.0} 
    \setlength{\tabcolsep}{5pt}
    \centering
\small
    \begin{tabular}{lcccc}
    \hline
        Operators & R$_y$ & CNOT & CR$_y$ & MCR$_y$\\
       \begin{tabular}{c}\end{tabular} & 
       \begin{tabular}{c}
            \Qcircuit @C=.3em @R=.5em {
                &&\\
                 & \gate{\text{R}_y} & \qw  \\
                &&\\
            }
        \end{tabular}
        &
       \begin{tabular}{c}
            \Qcircuit @C=.3em @R=.5em {
            &&\\
                & \ctrl{1} & \qw \\
                & \targ & \qw  \\
            }
        \end{tabular}
        &
       \begin{tabular}{c}
            \Qcircuit @C=.3em @R=.5em {
            &&\\
                & \ctrl{1} & \qw \\
                & \gate{\text{R}_y} & \qw 
            }
        \end{tabular}
        &
       \begin{tabular}{c}
            \Qcircuit @C=.3em @R=.5em {
             & \ctrl{1} & \qw \\
             & \ctrl{1} & \qw \\
             & \gate{\text{R}_y} & \qw  \\
            }
        \end{tabular}
        \vspace{.5em}\\
        \hline
        CNOT cost & 0 & 1 & 2 & $\mathcal{O}(2^n)$~\cite{mottonen2004quantum} \\
        \hline
    
    \end{tabular}
\end{table}

\subsection{Quantum Circuits for State Preparations}
Given a state $\ket{\psi}$ and a set of quantum gates $\mathcal{L}$, the \emph{quantum state preparation} finds a quantum circuit comprising $l$ gates $U_1, U_2, ... U_l\in \mathcal{L}$ such that these gates map the ground state $\ket{0}$ to the desired state $\ket{\psi}$, i.e., $\ket{\psi} = U_l...U_2U_1\ket{0}$. For \emph{noisy intermediate-scale quantum}~(NISQ) computers, CNOTs introduce more noise than single-qubit gates. Therefore, the objective of QSP is to minimize the CNOT cost of the circuit, which is the sum of the gates' CNOT costs.

\section{Motivating Example}\label{sec:motivation}
Consider the problem of preparing the state $\psi$, with $\ket{\psi} = \frac{1}{\sqrt{4}}(\ket{000} + \ket{011} + \ket{101} + \ket{110})$ which comprises three qubits with a cardinality of four. We demonstrate the quantum circuits generated by two categories of existing methods.

\begin{figure}[h]
    \centering
    \mbox{
        \small
        \Qcircuit @C=.7em @R=1em {
        \lstick{q_1: \ket{0}} & \rypi{2} & \ctrlo{1} & \ctrl{1}  & \ctrlo{1} & \ctrl{1}  & \qw  \gategroup{1}{3}{2}{4}{.5em}{--}\gategroup{1}{5}{3}{6}{.5em}{-}\\
        \lstick{q_2: \ket{0}} & \qw      & \rypi{2}  & \rypi{2}  & \ctrl{1} & \ctrlo{1} & \qw\\
        \lstick{q_3: \ket{0}} & \qw      &  \qw      &  \qw      & \ry{\pi} & \ry{\pi}  & \qw
        }
    }
    \caption{\small 6-CNOT circuit using the qubit reduction method.}\label{fig:qubit-reduction}
\vspace*{-3mm}
\end{figure}
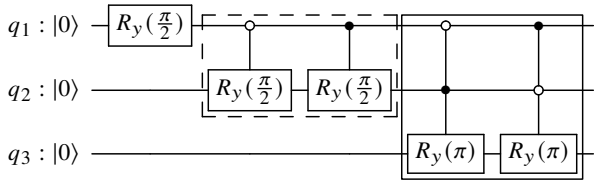

The first category of methods use \emph{qubit reduction}~\cite{mozafari2019preparation, araujo2021divide}. These methods focus on one target qubit at each stage and apply MCR$_y$ gates to separate it from the entanglement. As illustrated in \Cref{fig:qubit-reduction}, the gates in the solid box separate $q_3$, and the gates in the dashed box separate $q_2$. The dashed and solid boxes have 1 and 2 control qubits, respectively, and require $2^1\!+\!2^2\!=\!6$ CNOT gates. 

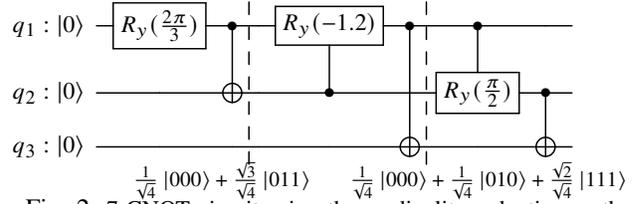
\begin{figure}[h]
    \centering
    \mbox
    {
        \small
        \Qcircuit @C=.7em @R=1em {
            \lstick{q_1: \ket{0}} & \gate{R_y(\frac{2\pi}{3})} & \ctrl{1} \ar@{--}[]+<.7em,1em>;[d]+<.7em,-3em> & \qw & \gate{R_y(-1.2)} & \ctrl{2}\ar@{--}[]+<.7em,1em>;[d]+<.7em,-3em>  & \ctrl{1} & \qw & \qw &\\
            \lstick{q_2: \ket{0}} & \qw & \targ  & \qw  & \ctrl{-1} & \qw   & \gate{R_y(\frac{\pi}{2})} & \ctrl{1} & \qw &\\
            \lstick{q_3: \ket{0}} & \qw & \qw   & \qw  &  \qw & \targ   & \qw & \targ & \qw & \\
            && \mbox{\footnotesize $\frac{1}{\sqrt{4}}\ket{000} + \frac{\sqrt{3}}{\sqrt{4}}\ket{011}\quad$} &&&& \mbox{\footnotesize \quad$\frac{1}{\sqrt{4}}\ket{000} + \frac{1}{\sqrt{4}}\ket{010} + \frac{\sqrt{2}}{\sqrt{4}}\ket{111}$} \\
        }
    }
    \caption{\small 7-CNOT circuit using the cardinality reduction method.}\label{fig:cardinality-reduction}
\vspace*{-3mm}
\end{figure}

Other related works perform \emph{cardinality reduction}~\cite{gleinig2021efficient}. This method iteratively selects two different basis vectors from the index set and merges them using CNOT and controlled rotation gates. As depicted in \Cref{fig:cardinality-reduction}, the circuit strictly reduces cardinality by one after each ``merging'' from right to left until $|S|=1$, which is the ground state, $\ket{000}$. This circuit contains a single-qubit gate, three CNOTs, and two controlled Y-rotation gates. Therefore, the CNOT cost is $3\!\times\!1\!+\!2\!\times\!2\!=\!7$.

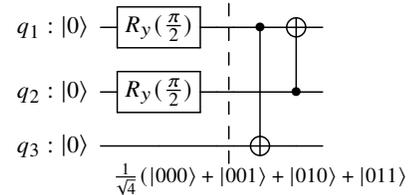
\begin{figure}[h]
    \centering
    \mbox
    {
        \small
        \Qcircuit @C=.7em @R=1em {
            \lstick{q_1: \ket{0}} & \gate{R_y(\frac{\pi}{2})} & \qw \ar@{--}[]+<.5em,1em>;[d]+<.5em,-3em>  & \qw & \ctrl{2} & \targ            & \qw \\
            \lstick{q_2: \ket{0}} & \gate{R_y(\frac{\pi}{2})} & \qw                                        & \qw & \qw      & \ctrl{-1}        & \qw \\
            \lstick{q_3: \ket{0}} & \qw                       & \qw                                        & \qw & \targ    & \qw              & \qw \\
            &&&& \mbox{\footnotesize $\frac{1}{\sqrt{4}}(\ket{000} + \ket{001} + \ket{010} + \ket{011}$} \\
        }
    }
    \caption{\small 2-CNOT circuit using the exact synthesis (ours).}\label{fig:exact-synthesis-circuit}
\vspace*{-3mm}
\end{figure}

Both qubit- and cardinality-reduction methods divide and conquer the QSP to decrease complexity. However, their heuristics introduce structural constraints to the circuit, limiting the visibility in the solution space. As a result, neither approach can reach the solution in \Cref{fig:exact-synthesis-circuit} that correctly prepares the state $\ket{\psi}$ using fewer CNOT gates. Indeed, qubit reduction methods target each qubit consecutively, while in \Cref{fig:exact-synthesis-circuit}, two gates (an $\text{R}_y$ and a CNOT) targeting $q_1$ are separated by a CNOT gate; similarly, each (controlled-) rotation gate in cardinality reduction methods strictly reduce the cardinality by one, while the two $\text{R}_y$ gates shrink the cardinality directly from four to one. 

This example points to the need to eliminate unnecessary constraints in the problem formulation, thereby reducing the number of CNOT gates required. More specifically, heuristics that rigidly guide the state transition process often lead to greedy decisions that are locally advantageous, such as reducing entanglement between qubits or decreasing cardinality, but are globally suboptimal. The following sections demonstrate our method that explores various directions for state transition to identify the optimal solution. 

\section{Exact CNOT Synthesis Formulation}\label{sec:formulation}
Based on the necessity to explore state transitions comprehensively, we formulate the QSP as a shortest path problem. In this section, we will first detail the problem formulation and then discuss the selection of the gate library, with an emphasis on the complexity analysis.

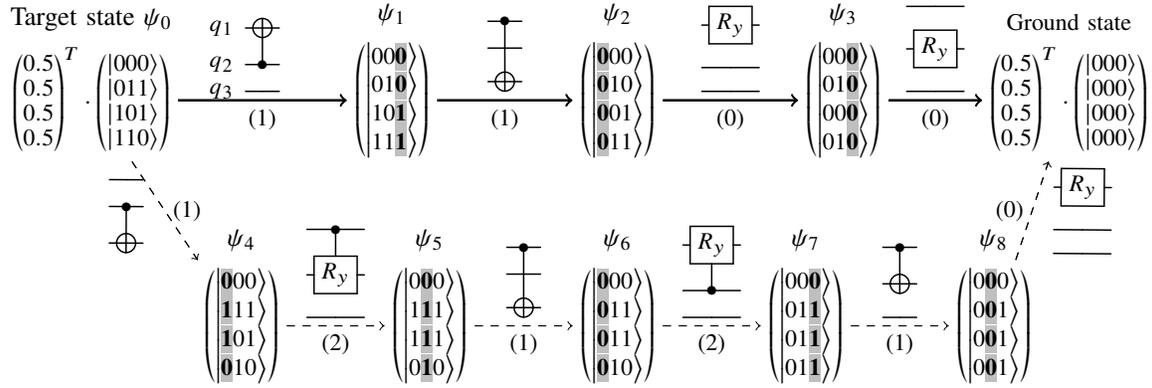
\begin{figure*}

\newcommand{\reducedstrut}{\vrule width 0pt height 1.15\ht\strutbox depth .9\dp\strutbox\relax}
\newcommand{\changed}[1]{\begingroup\fboxsep=0pt \colorbox{lightgray}{\reducedstrut \textcolor{black}{\textbf{#1}}}\endgroup}

\small
    \centering

    \begin{tikzpicture}
    
    \node[black, label={\normalsize Target state $\psi_0$}] (s0) at (0,0) {$
    \begin{pmatrix}
        0.5 \\
        0.5 \\
        0.5 \\
        0.5 \\
    \end{pmatrix}^T\cdot
    \begin{pmatrix}
        \ket{000} \\
        \ket{011} \\
        \ket{101} \\
        \ket{110} \\
    \end{pmatrix}
    $};

    \node[black, label={\normalsize $\psi_1$}] (s1) at (4,0) {$
    \begin{pmatrix}
        \ket{00\changed{0}} \\
        \ket{01\changed{0}} \\
        \ket{10\changed{1}} \\
        \ket{11\changed{1}} \\
    \end{pmatrix}
    $};
    \draw[solid, black, thick, ->] (s0) -- (s1) node [midway, above, sloped] (TextNode) {$
        \begin{array}{cc}
        \Qcircuit @C=.3em @R=1em {
            \lstick{q_1} & \targ & \qw  \\
            \lstick{q_2} & \ctrl{-1} & \qw  \\
            \lstick{q_3} & \qw & \qw  
        }
        \end{array}
    $};
    \draw[solid, black, thick, ->] (s0) -- (s1) node [midway, below, sloped] (TextNode) {(1)};

    \node[black, label={\normalsize $\psi_2$}] (s2) at (7,0) {$
    \begin{pmatrix}
        \ket{\changed{0}00} \\
        \ket{\changed{0}10} \\
        \ket{\changed{0}01} \\
        \ket{\changed{0}11} \\
    \end{pmatrix}
    $};
    \draw[solid, black, thick, ->] (s1) -- (s2) node [midway, above, sloped] (TextNode) {$
        \begin{array}{cc}
        \Qcircuit @C=.3em @R=1em {
            & \ctrl{2} & \qw  \\
            & \qw & \qw  \\
            & \targ & \qw 
        }
        \end{array}
    $};
    \draw[solid, black, thick, ->] (s1) -- (s2) node [midway, below, sloped] (TextNode) {(1)};

    \node[black, label={\normalsize $\psi_3$}] (s3) at (10,0) {$
    \begin{pmatrix}
        \ket{00\changed{0}} \\
        \ket{01\changed{0}} \\
        \ket{00\changed{0}} \\
        \ket{01\changed{0}} \\
    \end{pmatrix}
    $};
    \draw[solid, black, thick, ->] (s2) -- (s3) node [midway, above, sloped] (TextNode) {$
        \begin{array}{cc}
        \Qcircuit @C=.3em @R=1em {
          & \gate{R_y} & \qw  \\
          & \qw & \qw  \\
          & \qw & \qw  
        }
        \end{array}
    $};
    \draw[solid, black, thick, ->] (s2) -- (s3) node [midway, below, sloped] (TextNode) {(0)};

    \node[black, label = {Ground state}] (s4) at (13,0) {$
    \begin{pmatrix}
        0.5 \\
        0.5 \\
        0.5 \\
        0.5 \\
    \end{pmatrix}^T\cdot
    \begin{pmatrix}
        \ket{000} \\
        \ket{000} \\
        \ket{000} \\
        \ket{000} \\
    \end{pmatrix}
    $};
    \draw[solid, black, thick, ->] (s3) -- (s4) node [midway, above, sloped] (TextNode) {$
        \begin{array}{cc}
        \Qcircuit @C=.3em @R=1em {
             & \qw & \qw  \\
             & \gate{R_y} & \qw  \\
             & \qw & \qw 
        }
        \end{array}
    $};
    \draw[solid, black, thick, ->] (s3) -- (s4) node [midway, below, sloped] (TextNode) {(0)};

    \node[black, label={\normalsize $\psi_4$}] (s11) at (2,-3) {$
    \begin{pmatrix}
        \ket{\changed{0}00} \\
        \ket{\changed{1}11} \\
        \ket{\changed{1}01} \\
        \ket{\changed{0}10} \\
    \end{pmatrix}
    $};
    \draw[dashed, ->] (s0) -- (s11) node [midway, left, sloped, rotate=56] (TextNode) {$
        \begin{array}{cc}
        \Qcircuit @C=.3em @R=1em {
             & \qw & \qw  \\
             & \ctrl{1} & \qw  \\
             & \targ & \qw 
        }
        \end{array}
    $};
    \draw[dashed] (s0) -- (s11) node [midway, right, sloped, rotate=56] (TextNode) {(1)};

    \node[black, label={\normalsize $\psi_5$}] (s12) at (4.5,-3) {$
    \begin{pmatrix}
        \ket{0\changed{0}0} \\
        \ket{1\changed{1}1} \\
        \ket{1\changed{1}1} \\
        \ket{0\changed{1}0} \\
    \end{pmatrix}
    $};
    \draw[dashed, ->] (s11) -- (s12) node [midway, above, sloped] (TextNode) {$
        \begin{array}{cc}
        \Qcircuit @C=.3em @R=1em {
             & \ctrl{1} & \qw  \\
             & \gate{R_y} & \qw  \\
             & \qw & \qw 
        }
        \end{array}
    $};
    \draw[dashed] (s11) -- (s12) node [midway, below, sloped] (TextNode) {(2)};

    \node[black, label={\normalsize $\psi_6$}] (s13) at (7,-3) {$
    \begin{pmatrix}
        \ket{\changed{0}00} \\
        \ket{\changed{0}11} \\
        \ket{\changed{0}11} \\
        \ket{\changed{0}10} \\
    \end{pmatrix}
    $};
    \draw[dashed, ->] (s12) -- (s13) node [midway, above, sloped] (TextNode) {$
        \begin{array}{cc}
        \Qcircuit @C=.3em @R=1em {
             & \ctrl{2} & \qw  \\
             & \qw & \qw \\
             & \targ & \qw  
        }
        \end{array}
    $};
    \draw[dashed] (s12) -- (s13) node [midway, below, sloped] (TextNode) {(1)};

    \node[black, label={\normalsize $\psi_7$}] (s14) at (9.5,-3) {$
    \begin{pmatrix}
        \ket{00\changed{0}} \\
        \ket{01\changed{1}} \\
        \ket{01\changed{1}} \\
        \ket{01\changed{1}} \\
    \end{pmatrix}
    $};
    \draw[dashed, ->] (s13) -- (s14) node [midway, above, sloped] (TextNode) {$
        \begin{array}{cc}
        \Qcircuit @C=.3em @R=1em {
             & \gate{R_y} & \qw  \\
             & \ctrl{-1} & \qw \\
             & \qw & \qw  
        }
        \end{array}
    $};
    \draw[dashed] (s13) -- (s14) node [midway, below, sloped] (TextNode) {(2)};

    \node[black, label={\normalsize $\psi_8$}] (s15) at (12,-3) {$
    \begin{pmatrix}
        \ket{0\changed{0}0} \\
        \ket{0\changed{0}1} \\
        \ket{0\changed{0}1} \\
        \ket{0\changed{0}1} \\
    \end{pmatrix}
    $};
    \draw[dashed, ->] (s14) -- (s15) node [midway, above, sloped] (TextNode) {$
        \begin{array}{cc}
        \Qcircuit @C=.3em @R=1em {
             & \ctrl{1} & \qw  \\
             & \targ & \qw \\
             & \qw & \qw  
        }
        \end{array}
    $};
    \draw[dashed] (s14) -- (s15) node [midway, below, sloped] (TextNode) {(1)};

    \draw[dashed, ->] (s15) -- (s4) node [midway, right, sloped, rotate = -72] (TextNode) {$
        \begin{array}{cc}
        \Qcircuit @C=.3em @R=1em {
             & \gate{R_y} & \qw  \\
             & \qw & \qw \\
             & \qw & \qw  
        }
        \end{array}
    $};
    \draw[dashed] (s15) -- (s4) node [midway, left, sloped, rotate = -72] (TextNode) {(0)};

    \end{tikzpicture}
    
    \caption{\small Example of exploring a (portion of a) state transition graph from the target state $\ket{\psi} = \frac{1}{\sqrt{4}}(\ket{000} + \ket{011} + \ket{101} + \ket{110})$.}
    \label{fig:state-transition-graph}
\vspace*{-5mm}
\end{figure*}

\subsection{Shortest Path Problem Formulation}
Given a set of quantum gates $\mathcal{L}$ with fixed CNOT costs, we define \emph{state transition graph}, $G_{\mathcal{L}} = (V_G, A_G)$, where vertices are quantum states and arcs represent state transitions implementable by gates in $\mathcal{L}$. \Cref{tab:variables} lists the variables utilized in our formulation. An arc $a\!=\!(\psi, \varphi)$ in $A_G$ corresponds to an operator $U_a$ such that $\varphi = U_a \psi$. We define \emph{distance} of $a$, denoted by $d(a)$, as the given CNOT cost of $U_a$. 
\begin{table}[b]
\renewcommand{\arraystretch}{1.2} 
    \caption{\small Variable declarations for the shortest path problem.}
    \label{tab:variables}
\small
    \centering
    \begin{tabular}{ll}
    \hline
        $d(a)$ & The distance on an arc $a$, $a \in A_G$. \\
        $\gamma(\psi, \varphi)$ & The CNOT cost to prepare state $\psi$ from $\varphi$. \\
        $\hat\delta(\psi, \varphi)$ & The distance estimation between $\psi$ and $\varphi$. \\
    \hline
    \end{tabular}
\end{table}

We use $\gamma(\psi, \varphi)$ to represent the CNOT cost to prepare $\varphi$ from the state $\psi$. For a given target state $\psi_0$, our objective is to find the \emph{optimal} circuit that minimizes $\gamma(\ket{0}, \psi_0)$.

As an example, \Cref{fig:state-transition-graph} displays a portion of the state transition from the example in \Cref{sec:motivation}. The edges in \Cref{fig:state-transition-graph} represent quantum gates that accomplish the state transition. We label the gates and the corresponding CNOT costs on the graph. The optimal solution in \Cref{fig:exact-synthesis-circuit} is highlighted in bold lines, while the suboptimal \Cref{fig:cardinality-reduction} corresponds to the dashed path. Once the state transition graph is available, we can accurately evaluate and choose between these two solutions: the bold path has a distance of $1\!+\!1\!=\!2$, while the distance of the dashed path is $1\!+\!2\!+\!1\!+\!2\!+\!1\!=\!7$.



\subsection{Specialized Set of Quantum Gates for QSP}\label{subsec:library}
The selection of library $\mathcal{L}$ determines the graph size since the vertices $V_G$ include all reachable states using gates in $\mathcal{L}$. Let $\epsilon$ be the precision level of amplitudes. A universal set of quantum gates such as $\{\text{CNOT}, \mathcal{U}(2)\}$ can reach all $n$-qubit states ($\approx\epsilon^{-2^n}$). As $\epsilon$ approaches zero, the graph size increases dramatically to infinity, and the shortest path problem is impractical to solve. Therefore, we introduce a specialized set of quantum gates for QSP to upper bound the complexity.

We define \emph{amplitude-preserving}~(AP) mapping between quantum states as a transition where amplitude values are conserved, but the associated basis vectors, or indices, are changed. State transitions between $\psi$ and $\varphi$ is AP if we can express them as $\psi=\sum_xc_x\ket{x}$ and $\varphi=\sum_xc_x\ket{f(x)}$, where $\ket{x}$ and $\ket{f(x)}$ are basis vectors and $c_x$ is the amplitude. When multiple indices map to the same index, i.e., $y\!=\!f(x_1)\!=\!f(x_2)$, we merge their probability to maintain the quantum state's integrity, $c_y=\sqrt{c_{x_1}^2+c_{x_2}^2}$. We define our library, $\mathcal{L}_{\text{QSP}}$, as the set of all single-target amplitude-preserving state transitions. 

Note that all state transitions in \Cref{fig:state-transition-graph} are amplitude-preserving. The vertices, $\psi_1, \psi_{2}, ..., \psi_{8}$, have the amplitude vector of $(\sqrt{0.25},\sqrt{0.25},\sqrt{0.25},\sqrt{0.25})$ and differ only by their basis vectors. $\psi_6$, for instance, corresponds to the vector $(\ket{000}, \ket{011}, \ket{011}, \ket{010})$ with duplicated indices $\ket{011}$. It represents $\psi_6\!=\!\sqrt{0.25}\ket{000}\!+\!\sqrt{0.5}\ket{011}\!+\!\sqrt{0.25}\ket{010}$ after merging the two amplitudes of $\ket{011}$. 



\section{Shortest Path Algorithm}\label{sec:algorithm}
With our exact CNOT synthesis formulation, the problem of optimizing CNOT cost for quantum state preparation is equivalent to finding the shortest path between two vertices in the state transition graph. In this section, we introduce our algorithm to solve the shortest path problem and accelerate without any loss in optimality.

\subsection{Admissible Heuristic Function and A* Algorithm}\label{subsec:a-star}
A* algorithm is the general solution to the shortest path problem, whose pseudocode is shown in \Cref{alg:bfs}. A \texttt{while} loop at line $4$ traverses all the reachable quantum states in ascending order of CNOT cost from the target state $\psi$. We iteratively pop the top element in the queue, enumerate all of its adjacent states, and add those states to the queue if a lower distance is found. After reaching the ground state, our algorithm backtracks the path and returns the quantum circuit by mapping each edge to the corresponding quantum operator.

The key idea of the A* algorithm is to find a \emph{lower bound estimation} for the distance between state $\psi$ and the destination, denoted by $\hat\delta(\psi, \ket{0})$. Then, the queue sorts the states based on the sum of $\gamma(\psi_0, \psi) + \hat\delta(\psi, \ket{0})$. This way, we prioritize the state more likely to be the shortest path that potentially reaches the destination and returns earlier. If the function is \emph{admissible}, i.e., $\hat\delta(\psi, \ket{0})$ always underestimates the true cost between $\varphi$ and $\ket{0}$, then the A* heuristic can prune unpromising branches and improve efficiency without loss in optimality. 

\begin{algorithm}[t]
\caption{A* algorithm}\label{alg:bfs}
\SetKwInOut{Input}{input}
\SetKwInOut{Output}{output}

 \DontPrintSemicolon
 \Input{The target quantum state $\psi_0$.}
 \Output{A sequence of quantum operators to prepare $\psi_0$ from the ground state.}
 
q $\gets \text{PriorityQueue()}$ \;
$\gamma(\psi_0, :) \gets \infty\,,\,\gamma(\psi_0, \psi_0) \gets 0$ \;
 q.put($\psi_0$) \;
\While{$q$ is not empty} {
    $\psi$ $\gets$ q.pop()\;
    \If{$\psi = \ket{0}$} {
        \textbf{break}\;
    }
    \For{$(\psi, \varphi) \in A_G$} {
        $\gamma'$ $\gets $ $\gamma(\psi_0, \Pi(\psi))$ + $d(\psi, \varphi)$.\;
        \If {$\gamma' \geq \gamma(\psi_0, \Pi(\varphi))$} {
            \textbf{continue}\;
        }
        $\gamma(\psi_0, \Pi(\varphi))$ $\gets$ $\gamma'$\;
        q.put($\varphi$)\;
    }
}
path $\gets$ back trace the edges from $\psi_0$ to ground state.\;
\Return path.\;
\end{algorithm}

Given a state $\psi$, we derive such a lower bound on CNOT by checking the number of entangled qubit pairs, which can be acquired by evaluating \emph{mutual information}~\cite{shannon1948mathematical}. Take state $\psi_1$ as an example. The \emph{cofactors} of $q_2$, which are the sets of indices with $q_2\!\!=\!\!1$ ($\left.\psi_1\right|_{q_2=1}$) and $q_2\!\!=\!\!0$ ($\left.\psi_1\right|_{q_2=0}$) are identical. 
\begin{equation*}
S\left(\left.\psi_1\right|_{q_2=1}\right) = S\left(\left.\psi_1\right|_{q_2=0}\right) = \left\{
    \ket{00}_{q_1q_3},
    \ket{11}_{q_1q_3}\right\},
\end{equation*}
which indicates $q_2$ might be separable. Meanwhile, the cofactors of $q_1$ and $q_3$ differ, implying a pair of entangled qubits. 


Note that the admissible heuristic function may underestimate the shortest distance of $\gamma(\ket{0}, \psi)$. For example, the minimum number of CNOT to prepare the 4-qubit GHZ state, $\ket{\psi} = \frac{1}{\sqrt{2}}(\ket{0000}+\ket{1111})$, is $3$, while the number of entangled qubits is $4$ and the heuristic returns $\lceil\frac{4}{2}\rceil=2$. 

\subsection{State Compression using Canonicalization}\label{subsec:canonicalization}
Using the A* algorithm, our solver does not explicitly construct the graph comprising all $2^{nm}$ states but explores a small portion around the given target. To further decrease the time complexity, we introduce a state compression heuristic that reduces the number of enqueued states.

The idea is based on \emph{canonicalization}. We associate states with their \emph{equivalence class}, denoted by $V_G / \sim$, a set of equivalent states under relation $\sim$, which is defined as follows:
\begin{itemize}
    \item Single-qubit unitaries, $\mathcal{U}(2)$. Two states $\psi$ and $\varphi$ are equivalent if $U\in\mathcal{U}(2)$ exists such that $\psi = U\varphi$. 
    \item Qubit permutation, designated by $\mathcal{P}$. Two states are equivalent if we swap the index of two qubits.
\end{itemize}

For example, the state $\varphi\!=\!\frac{1}{\sqrt{2}}(\ket{100}\!+\!\ket{010})$ is equivalent to:
\begin{itemize}
    \item $\psi_1\!=\!\frac{1}{\sqrt{2}}(\ket{000}\!+\!\ket{110})$ because a Pauli-X operation on the first or the second qubit would transition $\varphi$ to $\psi_1$.
    \item $\psi_2\!=\!\frac{1}{\sqrt{4}}(\ket{000}\!+\!\ket{001}\!+\!\ket{110}\!+\!\ket{111})$ because a $\text{R}_y(\frac{\pi}{2})$ gate on the last qubit transition $\varphi$ to $\psi_2$. 
    \item $\psi_3\!=\!\frac{1}{\sqrt{2}}(\ket{100}\!+\!\ket{001})$ because they are equivalent after swapping the definition of the second and the last qubit. 
\end{itemize}

We can utilize equivalent relations to compress states. Since the gates in $\mathcal{U}(2)$ have zero CNOT cost, the states $\psi, \varphi \in V_G / \mathcal{U}(2)$ can be prepared using the same number of CNOT gates. Similarly, if all qubits are interchangeable (which implies a symmetric coupling graph), then states $\psi, \varphi \in V_G / \mathcal{P}\mathcal{U}(2)$ should have the same CNOT cost. Therefore, we can select a \emph{representative} for each equivalence class to store the distance. As shown in lines 10 to 13 in \Cref{alg:bfs}, where $\Pi(\varphi)$ denotes the representative of $\varphi$, we can skip exploring state $\varphi$ if another state in the same equivalence class is enqueued with a lower or equal distance. 

\begin{table}[ht]
\renewcommand{\arraystretch}{1.0} 

\centering
\caption{\small Number of canonical $4$-qubit uniform states.}
\label{tab:canon4}
\begin{tabular}{r|r|r|r}
\hline
$m$ & $|V_G|$   & $|V_G / \mathcal{U}(2)|$  & $|V_G / \mathcal{P}\mathcal{U}(2)|$ \\
\hline
1        & 16    & 1   & 1   \\
2        & 120   & 11  & 3   \\
3        & 560   & 35  & 6   \\
4        & 1820  & 118 & 16  \\
5        & 4368  & 273 & 27  \\
6        & 8008  & 525 & 47  \\
7        & 11440 & 715 & 56  \\
8        & 12870 & 828 & 68  \\
\hline
\end{tabular}
\end{table}
\Cref{tab:canon4} presents the graph size without an equivalency relation ($|V_G|$), with a layout-variant equivalency $|V_G/\mathcal{U}(2)|$, and with a layout-invariant equivalency $|V_G/\mathcal{P}\mathcal{U}(2)|$. The results demonstrate the effectiveness of state compression using canonicalization.

\section{Evaluation}\label{sec:evaluation}
The proposed algorithm is implemented using Python and is open-source.\footnote{https://github.com/Nozidoali/quantum-xyz} In this section, we illustrate the effectiveness and efficiency of our method compared with manual designs and three recent works: a cardinality reduction method~(\textbf{``$m$-flow''})~\cite{gleinig2021efficient}, a qubit reduction method~(\textbf{``$n$-flow''})~\cite{mozafari2019preparation}, and the latest QSP solver: a hybrid method combining $n$- and $m$-flows~(\textbf{``hybrid''})~\cite{mozafari2022efficient}. The hybrid method requires one ancilla, while all others employ no ancilla qubits. We run all the experiments on a classical computer with a 3.7GHz AMD Ryzen 9 5900X processor with $64$GB of RAM. 

\begin{figure}[b]
    \centering
    \includegraphics[width=.75\linewidth]{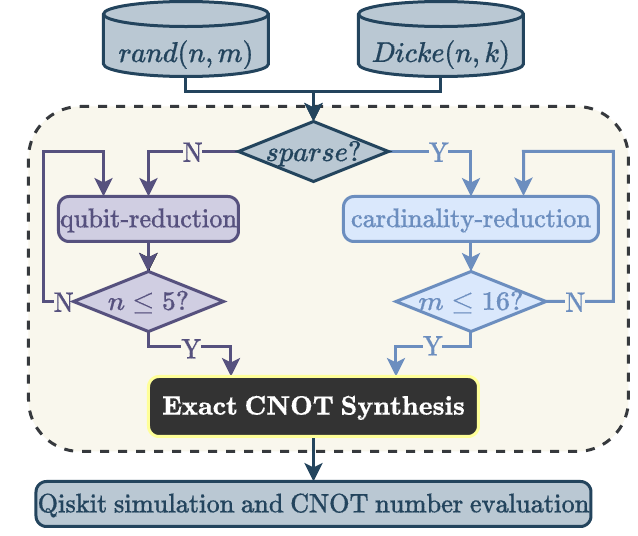}
    \caption{\small Evaluation workflow of exact CNOT synthesis.}
    \label{fig:workflow}
\end{figure}

\subsection{Workflow}\label{subsec:workflow}
The workflow is illustrated in \Cref{fig:workflow}, where we integrate our exact CNOT synthesis into a scalable framework. Given the state with $n$ qubits and cardinality $m$, we choose the divide-and-conquer method based on the sparsity. If the state is sparse ($nm < 2^n$), we iteratively run the cardinality reduction method until the complexity is acceptable for exact CNOT synthesis; otherwise, if the state is dense ($nm \geq 2^n$), we apply qubit reduction before an exact synthesis. We use the library introduced in \Cref{subsec:library}. We verify the correctness of the circuits returned by the QSP solver employing Qiskit simulators~\cite{Qiskit} and evaluate the number of CNOT gates after mapping the circuit to $\{\mathcal{U}(2), \text{CNOT}\}$. Although we test uniform states to compare with related works, our implementation applies to any state with real amplitudes. Employing a phase oracle, we can prepare arbitrary states with complex amplitudes~\cite{Amy_2018}. 

\subsection{Preparation of Dicke States}\label{subsec:results-dicke}
Dicke states, denoted by $\ket{D}_n^k$, are the family of quantum states of $n$ qubits with nonzero coefficient if $k$ out of the $n$ qubits in the basis state is $\ket{1}$. Due to its importance and wide applications~\cite{bartschi2019deterministic}, various manually designed techniques have been proposed~\cite{cruz2019efficient, bartschi2019deterministic, mukherjee2020preparing, aktar2022divide}. The latest manual design requires $(5nk-5k^2-2n)$ CNOT gates to prepare $\ket{D}_n^k$~\cite{mukherjee2020preparing}.

\begin{table}[t]
\caption{\small\label{tab:result-dicke}CNOT number comparison results of Dicke state preparation. We calculate the improvements achieved by four design automation algorithms compared to the manual design.}
\tabcolsep = 3.5pt
\renewcommand{\arraystretch}{1.0} 
\centering
\begin{tabular}{rr|rrrr|r}
\hline
\multicolumn{6}{c}{\textbf{Dicke states preparation$\ket{D}_n^k$}} \\
\hline
$n$ & $k$ & Manual~\cite{mukherjee2020preparing} & $m$-flow~\cite{gleinig2021efficient} & $n$-flow~\cite{mozafari2019preparation} & Hybrid~\cite{mozafari2022efficient} & \textbf{ours} \\
\hline
3 & 1 & \textbf{4}  & 5   & 6  & 27  & \textbf{4}  \\
4 & 1 & \textbf{7}  & 9   & 14 & 141 & \textbf{7}  \\
4 & 2 & 12          & 24  & 14 & 141 & \textbf{6}  \\
5 & 1 & \textbf{10} & 13  & 30 & 237 & \textbf{10} \\
5 & 2 & 20          & 67  & 30 & 601 & \textbf{16} \\
6 & 1 & \textbf{13} & 17  & 62 & 241 & \textbf{13} \\
6 & 2 & 28          & 117 & 62 & 645 & \textbf{22} \\
6 & 3 & 33          & 231 & 62 & 779 & \textbf{25} \\
\hline
\multicolumn{2}{l|}{\textbf{geo. mean}}    & 13.0   & 28.5   & 26.6  & 251.1  & \textbf{10.9}  \\
\hline
\multicolumn{2}{l|}{Impr\%}    & -    & -119\%     & -105\% & -1832\%  & \textbf{17\%}  \\
\hline
\end{tabular}
\vspace*{-4mm}
\end{table}

\Cref{tab:result-dicke} presents the CNOT numbers of each method and the improvements compared to the manual design~\cite{mukherjee2020preparing}, which demonstrates the effectiveness of the exact synthesis method on small states: our method achieves the best CNOT number among four methods on \textbf{all} the benchmarks. 

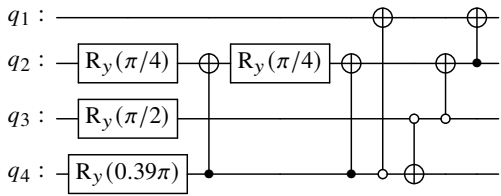
\begin{figure}[b]
\centering
\mbox{
    \small
    \Qcircuit @C=.5em @R=.7em {
        \lstick{q_1:} & \qw                       & \qw       & \qw                    & \qw       & \targ{}      & \qw & \qw & \targ{} & \qw \\
        \lstick{q_2:} & \gate{\text{$\text{R}_y$}(\pi/4)} & \targ{}   & \gate{\text{$\text{R}_y$}(\pi/4)}  & \targ{}   & \qw  & \qw  & \targ{} & \ctrl{-1} & \qw \\
        \lstick{q_3:} & \gate{\text{$\text{R}_y$}(\pi/2)} & \qw       & \qw                      & \qw       & \qw  & \ctrlo{1} & \ctrlo{-1} & \qw & \qw  \\
        \lstick{q_4:} & \gate{\text{$\text{R}_y$}(0.39\pi)} & \ctrl{-2} & \qw                    & \ctrl{-2} & \ctrlo{-3} & \targ{} & \qw & \qw  & \qw
    }
}
\caption{\small Circuit to prepare $\ket{D}_4^2$ using $6$ CNOTs}\label{fig:exact-synthesis-dicke-state}
\end{figure}

Moreover, our method is the first design automation algorithm that outperforms manual designs. Utilizing the exact CNOT synthesis formulation, our algorithm implicitly derives more complicated properties throughout the vast solution space exploration. While the properties developed manually are mostly symmetric or inductive, our solver synthesizes irregular yet effective circuits that are not easily generalizable, as shown in \Cref{fig:exact-synthesis-dicke-state}. We reduce the CNOT number from $12$ to $6$.

\subsection{Preparation of General Quantum States}\label{subsec:results-general}
Then, we evaluate the generality of our method by applying exact CNOT synthesis on asymmetry states with various numbers of qubits with different cardinalities. Our benchmark suites contain dense states with cardinality $m = 2^{n-1}$ and the sparse states with $m=n$. We randomly sample $100$ different states for each parameter setup and present the average number of CNOT gates in \Cref{tab:results-general}. 
\begin{table}[t]
\tabcolsep = 3.5pt
\caption{\small CNOT number comparison results of general state preparation. ``TLE'' represents that the method fails to prepare the state within a one-hour time limit. We calculate the improvement of our method compared to the most effective baseline in each category: to $n$-flow for dense states and to $m$-flow for sparse states.}
\label{tab:results-general}
\centering
\begin{tabular}{rr|rrr|rr}
\hline
\multicolumn{6}{c}{\textbf{Dense states preparation ($m = 2^{n-1}$)}} \\
\hline
$n$ & $m$ & $m$-flow~\cite{gleinig2021efficient} & $n$-flow~\cite{mozafari2019preparation} & Hybrid~\cite{mozafari2022efficient} & \textbf{ours} & impr\% \\
\hline
3  & 4      & 8       & 6       & 5                    & 5      & 17\% \\
4  & 8      & 40      & 14      & 28                   & 9      & 36\% \\
5  & 16     & 123     & 30      & 401                  & 29     & 3\%  \\
6  & 32     & 478     & 62      & 899                  & 56     & 10\% \\
7  & 64     & 1382    & 126     & 1268                 & 112    & 11\% \\
8  & 128    & 3954    & 254     & 2804                 & 226    & 11\% \\
9  & 256    & 10902   & 510     & 6911                 & 484    & 5\%  \\
10 & 512    & 28743   & 1022    & 15646                & 962    & 6\%  \\
11 & 1024   & 72441   & 2046    & 35650                & 1812   & 11\% \\
12 & 2048   & 178996  & 4094    & 82836                & 3846   & 6\%  \\
13 & 4096   & 440843  & 8190    & 183556               & 7746   & 5\%  \\
14 & 8192   & 1053633 & 16382   & 398602               & 15630  & 5\%  \\
15 & 16384  & 2487775 & 32766   & 879236               & 31254  & 5\%  \\
16 & 32768  & 5810670 & 65534   & 1915109              & 63720  & 3\%  \\
17 & 65536  & TLE     & 131070  & 4109698              & 128330 & 2\%  \\
18 & 131072 & TLE     & 262142  & 8802090              & 261684 & 0\%  \\
\hline
\multicolumn{2}{l}{\textbf{geo. mean}} & \multicolumn{1}{l}{}  & 1399.3 & 18855.9 & \textbf{1274.7} & \textbf{9\%} \\
\hline
\hline
\multicolumn{6}{c}{\textbf{Sparse states preparation ($m = n$)}} \\
\hline
$n$ & $m$ & $m$-flow~\cite{gleinig2021efficient} & $n$-flow~\cite{mozafari2019preparation} & Hybrid~\cite{mozafari2022efficient} & \textbf{ours} & impr\% \\
\hline
3                    & 3                    & 4    & 6       & 7     & 3   & 37\% \\
4                    & 4                    & 9    & 14      & 38    & 6   & 34\% \\
5                    & 5                    & 14   & 30      & 83    & 9   & 36\% \\
6                    & 6                    & 22   & 62      & 181   & 14  & 36\% \\
7                    & 7                    & 30   & 126     & 253   & 20  & 33\% \\
8                    & 8                    & 39   & 254     & 398   & 27  & 30\% \\
9                    & 9                    & 51   & 510     & 472   & 37  & 29\% \\
10                   & 10                   & 66   & 1022    & 560   & 44  & 33\% \\
11                   & 11                   & 80   & 2046    & 679   & 54  & 33\% \\
12                   & 12                   & 97   & 4094    & 763   & 66  & 32\% \\
13                   & 13                   & 114  & 8190    & 878   & 78  & 31\% \\
14                   & 14                   & 130  & 16382   & 965   & 91  & 30\% \\
15                   & 15                   & 152  & 32766   & 1095  & 106 & 30\% \\
16                   & 16                   & 172  & 65534   & 1187  & 119 & 31\% \\
17                   & 17                   & 201  & 131070  & 1268  & 139 & 31\% \\
18                   & 18                   & 217  & 262142  & 1399  & 155 & 29\% \\
19                   & 19                   & 238  & 524286  & 1491  & 173 & 28\% \\
20                   & 20                   & 267  & 1048574 & 1605  & 192 & 28\% \\
\hline
\multicolumn{2}{l}{\textbf{geo. mean}}  & 64.3 & 2809.3  & 429.8 & \textbf{44}  & \textbf{32\%} \\
\hline
\end{tabular}
\vspace*{-5mm}
\end{table}

We observe that the disparity in divide-and-conquer approaches between $n$-flow and $m$-flow leads to distinct benefits. When preparing states with $n$ qubits and cardinality of $m$, the $n$-flow exhibits a CNOT count bounded by $\mathcal{O}(2^n)$, which is well-suited for dense states. On the other hand, the $m$-flow requires a CNOT count of $\mathcal{O}(mn)$ and proves advantageous for sparse states. Meanwhile, the CNOT number of the hybrid method falls between the $n$-flow and the $m$-flow.

After employing the exact CNOT synthesis, we improve the results on both dense and sparse states. Compared with the corresponding advantageous baseline, our method reduces the CNOT number by $9\%$ and $32\%$ on dense and sparse states, respectively. Since we set fixed thresholds ($n\leq 4$ and $m\leq 16$) to activate the exact synthesis in our workflow, the room for improving the CNOT number does not scale with the problem size. Therefore, the relative benefit of exact synthesis decreases as $n$ and $m$ increases. We are investigating a more effective integration of the exact synthesis to scale the improvements.

\subsection{CPU Time Analysis}
\begin{figure}[t]
\captionsetup[subfigure]{justification=centering}
    \centering
    \begin{subfigure}[b]{.48\linewidth}
    \centering
        \includegraphics[width=.99\linewidth]{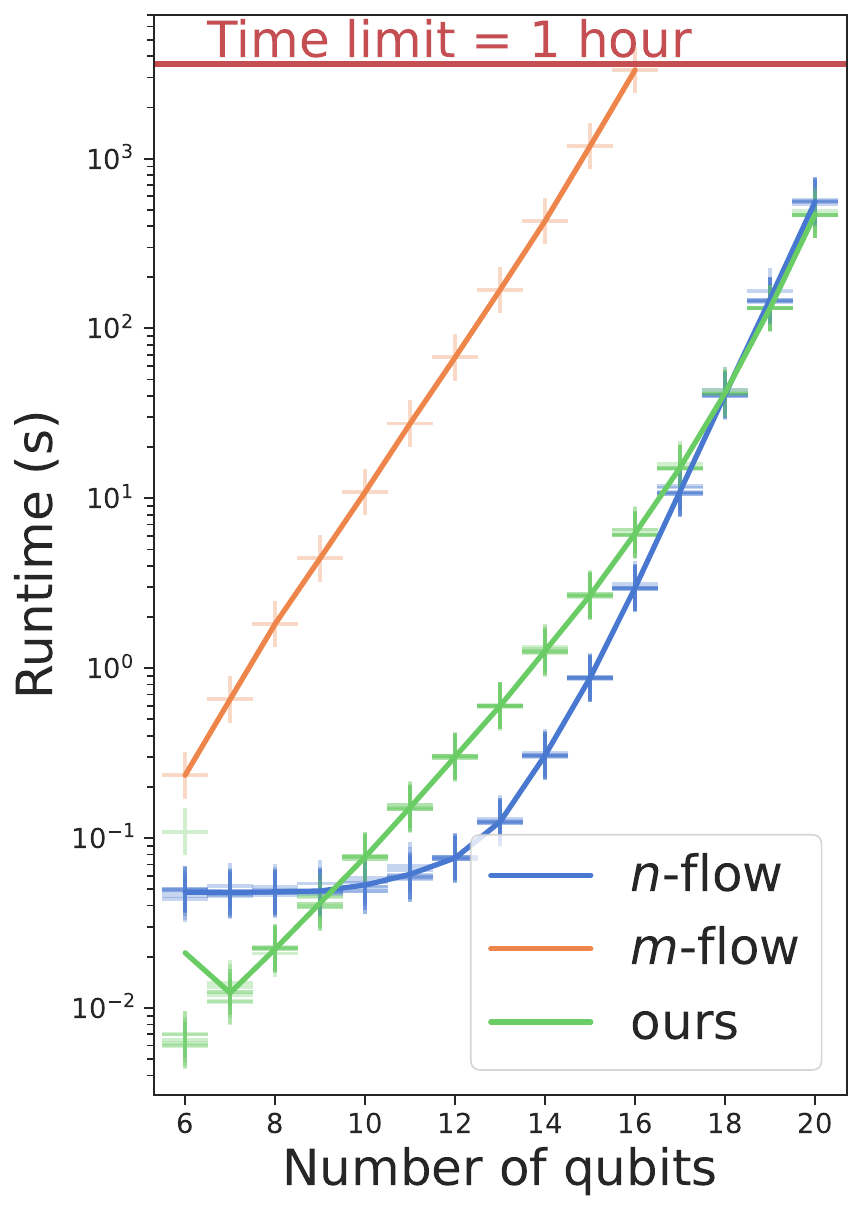}
        \caption{Dense states ($m=2^{n-1}$)}
        \label{fig:results-runtime-dense}
    \end{subfigure}
    \begin{subfigure}[b]{.48\linewidth}
    \centering
        \includegraphics[width=.99\linewidth]{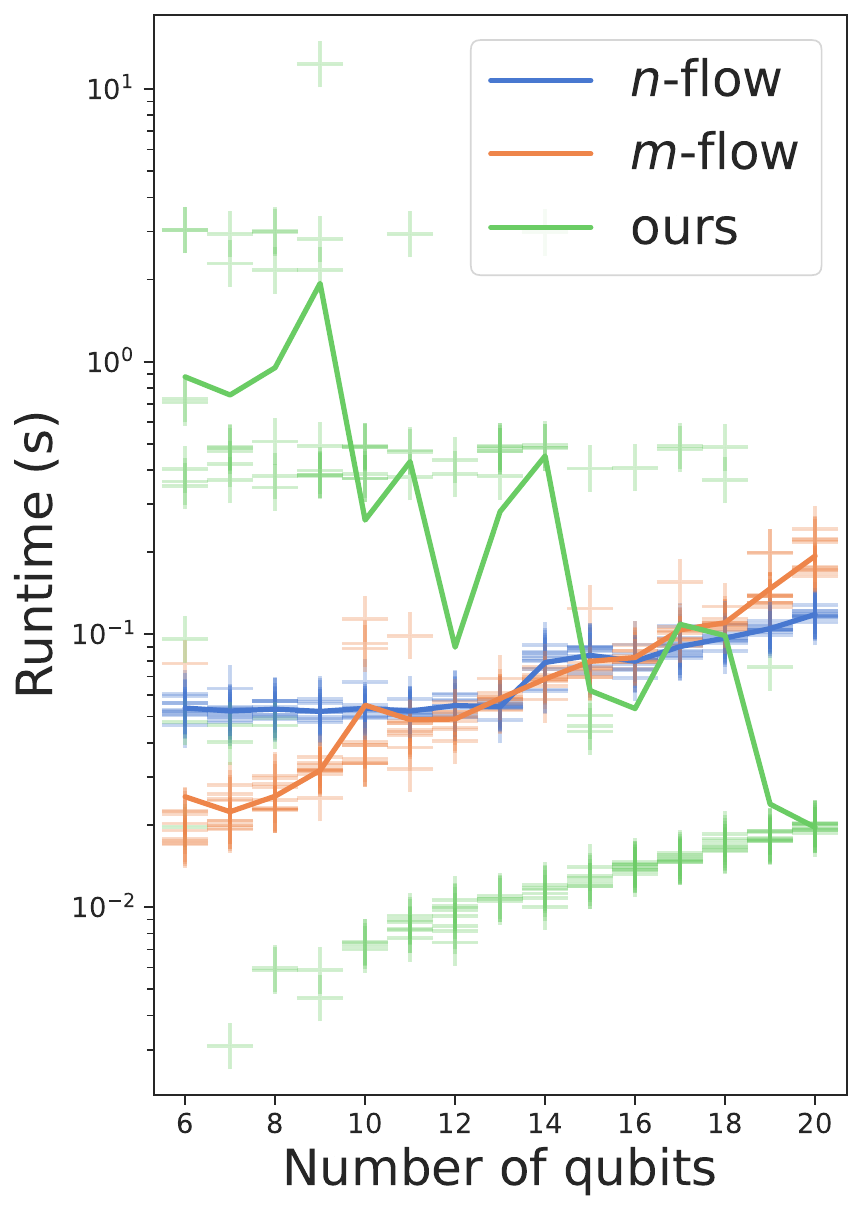}
        \caption{Sparse states ($m=n$)}
        \label{fig:results-runtime-sparse}
    \end{subfigure}
    \caption{\small CPU time analysis on (a) dense states and (b) sparse states compared with two baselines: the $n$-flow~\cite{mozafari2019preparation}, and the $m$-flow~\cite{gleinig2021efficient}. We omit the CPU time of the hybrid method~\cite{mozafari2022efficient}, which falls between $n$-flow and $m$-flow, for clarity.}
    \label{fig:cpu-time}
\vspace*{-5mm}
\end{figure}

Exact CNOT synthesis searches the shortest path with full visibility of the solution space, which naturally has a higher time complexity than the baselines. We investigate the scalability comparison in \Cref{fig:cpu-time}, where we plot the CPU time with the growth of qubit numbers. We show the results for dense states and sparse states separately in \Cref{fig:results-runtime-dense} and \Cref{fig:results-runtime-sparse}, as their divide-and-conquer procedures are different, as mentioned in \Cref{subsec:workflow}. 

The results demonstrate the robustness of our implementation. First, we use $n\times m$ classical bits to encode a quantum state while the $m$-flow~\cite{gleinig2021efficient} inherits Qiskit's data structure~\cite{Qiskit}, which stores indices as \texttt{string}. The $n$-flow~\cite{mozafari2019preparation} and the hybrid method~\cite{mozafari2022efficient} use \texttt{C++} CUDD\footnote{CUDD: CU Decision Diagram package, https://github.com/ivmai/cudd.} library with truth tables of $2^n$ bits. Second, our canonicalization method helps filter out separable qubits, further compressing the states and accelerating the solver (especially for sparse states). Therefore, although integrated with exact CNOT synthesis, our flow consumes comparable CPU time as baseline methods and exhibits better scalability as the number of qubits increases.

\section{Conclusion}
\emph{Quantum state preparation}~(QSP) initializes quantum superposition which is essential in quantum computing. Synthesizing efficient circuits for QSP improves the accuracy of quantum algorithms. However, the characterization of superposition and entanglement hinders the development of classical algorithms to automate quantum designs. This paper formulates QSP as a shortest path problem, encodes quantum states on a graph, and finds the optimal circuit with the lowest CNOT cost. Equipped with the A* algorithm and state compression heuristics, our method solves QSP efficiently without any loss in optimality. Compared to existing design automation algorithms, our method improves the CNOT number by $9\%$ and $32\%$ for dense and sparse state preparation, on average, using comparable CPU time. On a practical QSP problem, we reduce the best CNOT cost by \expnum{$2\times$}, which is the first time design automation algorithms surpass manual designs.


{
\footnotesize
\bibliographystyle{IEEEtran}
\bibliography{main}
}

\end{document}